\RequirePackage{fixltx2e}
\documentclass[aps,prb,reprint,floatfix]{revtex4-1}

\usepackage{amsmath,amsfonts,amssymb} %
\usepackage{mathtools} %
\usepackage{wasysym} %
\usepackage{bm} %
\usepackage[bbgreekl]{mathbbol} %
\usepackage{graphicx} %
\usepackage[dvipsnames,table]{xcolor} %
\usepackage{tabularx} %
\usepackage[acronym]{glossaries} %
\usepackage{braket} %
\usepackage{fancyhdr} %
\usepackage{microtype} %
\usepackage{natmove}
\usepackage[byname]{smartref} %
\usepackage[breaklinks, %
colorlinks, linkcolor=Blue, citecolor=Blue, urlcolor=Blue]{hyperref} %
\usepackage[version=3]{mhchem}
\AtBeginDocument{\usepackage{booktabs}} % Fix incompatibility with
\newcommand{\E}{\textrm{e}} %
\newcommand{\I}{\mathrm{i}\mkern1mu} %
\newcommand{\eq}[1]{Eq.~\eqref{#1}}
\def\be{\begin{equation}}
\def\ee{\end{equation}}
\def\bea{\begin{eqnarray}}
\def\eea{\end{eqnarray}}

\def\MyTitle{Relation between fermionic and qubit mean fields in the electronic structure problem} %
\def\MyAuthora{Ilya G. Ryabinkin} %
\def\MyAuthorb{Scott N. Genin} %
\def\MyAuthord{Artur F. Izmaylov} %
\def\MySubject{Quantum computing, quantum chemistry} %

\hypersetup{ %
  pdftitle={\MyTitle{}}, %
  pdfauthor={\MyAuthora{}, \MyAuthorb{}, and \MyAuthord{}}, %
  pdfsubject={\MySubject{}}, %
  pdfkeywords={} %
}

\newacronym{QPE}{QPE}{quantum phase estimation} %
\newacronym{VQE}{VQE}{variational quantum eigensolver} %
\newacronym{DFT}{DFT}{density functional theory} %
\newacronym{MSO}{MSO}{molecular spin orbital} 
\newacronym{FCI}{FCI}{full configurational interaction} %
\newacronym{JW}{JW}{Jordan--Wigner} %
\newacronym{BK}{BK}{Bravy--Kitaev} %
\newacronym[longplural={degrees of freedom}, %
firstplural={degrees of freedom (DOF)}, plural={DOF}]{DOF}{DOF}{degree
  of freedom} %
\newacronym[longplural={equations of motion}, %
firstplural={equations of motion (EOM)}, %
plural={EOM}]{EOM}{EOM}{equation of motion} %
\newacronym{PES}{PES}{potential energy surface} %
\newacronym{CI}{CI}{configuration interaction} %
\newacronym{QMF}{QMF}{qubit mean field} %
\newacronym{SQP}{SQP}{sequential quadratic programming} %
\newacronym{RHF}{RHF}{restricted Hartree--Fock}
\newacronym{HF}{HF}{Hartree--Fock}

\begin{document}

\title{\MyTitle}

\author{\MyAuthora{}} %
\affiliation{Department of Physical and Environmental Sciences,
  University of Toronto Scarborough, Toronto, Ontario\, M1C\,1A4,
  Canada}
\affiliation{Chemical Physics Theory Group, Department of Chemistry,
  University of Toronto, Toronto, Ontario\, M5S 3H6, Canada}

\author{\MyAuthorb{}} %
\affiliation{OTI Lumionics Inc., 100 College St. \#351, Toronto,
  Ontario\, M5G 1L5, Canada} %

% \author{\MyAuthorc{}} %
% \affiliation{Chemical Physics Theory Group, Department of Chemistry,
% University of Toronto, Toronto, Ontario M5S\,3H6, Canada} %

\author{\MyAuthord{}} %
\affiliation{Department of Physical and Environmental Sciences,
  University of Toronto Scarborough, Toronto, Ontario\, M1C 1A4,
  Canada}
\affiliation{Chemical Physics Theory Group, Department of Chemistry,
  University of Toronto, Toronto, Ontario\, M5S 3H6, Canada}

\date{\today}

\begin{abstract}
For quantum computing applications,  
the electronic Hamiltonian for the electronic structure problem needs to be 
unitarily transformed to a qubit form. 
We found that mean-field procedures on the original electronic Hamiltonian and on its transformed 
qubit counterpart can give different results. We establish conditions of when fermionic and qubit 
mean fields provide the same or different energies. In cases when the fermionic mean-field 
(Hartree--Fock) approach provides an accurate description 
(electronic correlation effects are small), the choice of molecular 
 orbitals for the electron Hamiltonian representation 
 becomes the determining factor in whether the qubit mean-field energy will be equal to or 
 higher than that of the fermionic counterpart. In strongly correlated cases, the qubit mean-field 
 approach has a higher chance to undergo symmetry breaking and lower its energy below the 
 fermionic counterpart.
\end{abstract}

\glsresetall

\maketitle

%%%%%%%%%%%%%%%%%%%%%%%%%%%%%%%%%%%%%%%%%%%%%%%%%%%%%%%%%%%%%%%%%%%%%%
\section{Introduction}

% 1) quantum computing grows - propaganda -> One of the difficulty is the coherence time 
% 2) VQE is one of the practical approaches - it starts w/ QMF

An interest in solving quantum chemistry problems on a quantum
computer has experienced a rapid growth in last two decades.\cite{QConQC} 
To address some of the technical difficulties (mainly maintaining long coherence times) 
associated with the initial \gls{QPE}\cite{Kitaev:1995/ArXiv/quant-ph/9511026,
  Abrams:1997/prl/2586, Abrams:1999/prl/5162,AspuruGuzik:2005/sci/1704,
  Brown:2006/prl/050504,Whitfield:2011/mp/735} approach, 
  the \gls{VQE}\cite{Peruzzo:2014/ncomm/4213, Jarrod:2016/njp/023023,
  OMalley:2016/prx/031007, Hempel:2018/ArXiv/1803.10238,Kandala:2017/nature/242} scheme has been 
  suggested recently and already proven to be feasible in calculating 
  \glspl{PES} of small molecules (e.g., H$_2$, LiH, and BeH$_2$). 
The \gls{VQE} approach searches for unitary rotations [$U_\text{ENT}(\boldsymbol \tau)$ 
and $U_\text{MF}(\boldsymbol \Omega)$] of qubits encoding the electronic wavefunction  
\bea
  \label{eq:ibm_vqe_ansatz}
    \ket{\Psi(\boldsymbol \Omega,\boldsymbol \tau)} &= & U_\text{ENT}(\boldsymbol \tau) U_\text{MF}(\boldsymbol \Omega) \ket{0_q},
\eea
that minimize the total electronic energy
\bea
  \label{eq:barE}
  E_e = \min_{\boldsymbol \Omega,\boldsymbol\tau} \braket{\Psi(\boldsymbol \Omega,\boldsymbol\tau)|
    H|\Psi(\boldsymbol \Omega,\boldsymbol\tau)},
\eea
where $\{\boldsymbol \Omega,\boldsymbol \tau\}$ are parameters of the unitary rotations, 
$\ket{0_q}$ is the initial state of qubits, and $H$ is the system electronic Hamiltonian. 
A typical starting point of this search is to introduce individual qubit rotations based on the results of a 
mean-field procedure, $U_\text{MF}(\boldsymbol \Omega)$. This \gls{QMF} is followed by the qubit 
entangling step, employing $U_\text{ENT}(\boldsymbol \tau)$. Evaluation of the expectation value in 
\eq{eq:barE} can be exponentially hard on a classical computer due to the entangling 
transformation $U_\text{ENT}$, whereas it has a polynomial scaling on a universal quantum computer.
Therefore, \gls{VQE} employs a hybrid scheme where a quantum computer evaluates the energy 
estimate for a particular set of parameters $\{\boldsymbol \Omega,\boldsymbol \tau\}$, and then, 
a classical computer finds the energy minimum.  
For calculating the energy expectation value, the Hamiltonian needs to be transformed from its 
original second-quantized fermionic form to a form operating in the qubit space. 
This fermion-to-qubit transformation is achieved by either the
\gls{JW}~\cite{Jordan:1928/zphys/631, AspuruGuzik:2005/sci/1704} or 
the \gls{BK}~\cite{Bravyi:2002/aph/210, Seeley:2012/jcp/224109,
  Tranter:2015/ijqc/1431} transformation. %Interestingly, depending on  

Although the entangling transformation can be generated using the Unitary Coupled Clusters 
(UCC)\cite{Peruzzo:2014/ncomm/4213}
approach, which provides energies at the level of chemical accuracy, 
practical truncated UCC procedures do not guarantee solutions equivalent to 
the exact Full Configuration Interaction (FCI) solution of the problem. 
In this case, any practical and approximate 
entangling scheme will depend on the quality of the mean-field result serving as the first step. 

This work is focused on differences in \gls{QMF} outcomes that originate from  
representations of the initial electronic Hamiltonian that use different one-particle basis functions.
Interestingly, results of the \gls{QMF} approach are not invariant with respect to different choices 
of one-electron basis functions. Here, we investigate the origins of this dependence and relations 
between results of fermionic (Hartree--Fock) and \gls{QMF} mean fields.  

The rest of the paper is organized as follows. First, we briefly recapitulate the
relevant basics of the \gls{HF} method and formally introduce the
\gls{QMF} method. Second, we illustrate the problem by considering 
different \gls{QMF} solutions for the \ce{LiH} molecule. 
Third, we discuss a simple one-electron Hamiltonian that can be treated exactly at the
\gls{HF} level and see what solutions \gls{QMF} can provide. 
Finally, we consider a strongly correlated case of a stretched \ce{H_2} molecule and discuss the 
symmetry breaking phenomenon. In conclusions, we summarize 
all the findings and discuss some of the perils of the \gls{QMF} method.

\section{Theory}
\label{sec:theory}

\subsection{Fermionic mean field}
\label{sec:hartr-fock-ferm}

An $N_e$-electron wavefunction in the \gls{HF} theory is obtained by
application of an ordered string of $N_e$ fermionic creation operators 
$a^\dagger_i$ to the mathematical vacuum state, $\ket{0}$, 
the state with zero electrons:
\begin{equation}
  \label{eq:wf_anz}
  \ket{\Phi} = \prod_{i=1}^{N_e} a^\dagger_i \ket{0}.
\end{equation}
Each creation operator $a^\dagger_i$ adds to a system an electron in a
state given by a single-particle \gls{MSO}, $\psi^{*}_i(\mathbf{x})$. Here, $\mathbf{x} = (\mathbf{r}, \sigma)$, 
  is a joined spatial and spin coordinate of an electron. For simplicity, we will consider only the 
  restricted version of the \gls{HF} theory, which implies that for every spatial single-particle function 
  there are two \glspl{MSO} with the opposite spins.  

Individual fermionic operators can be rotated using a unitary transformation, 
which conserves the fermionic anti-commutation relations,
\begin{equation}
  \label{eq:Ua}
  {\tilde a}^\dagger_i(\boldsymbol \kappa) = \exp(\hat \kappa)a^\dagger_i\exp(-\hat \kappa), 
\end{equation}
where 
\begin{equation}
  \label{eq:kappa}
  \hat \kappa  = \sum_{i>j} \kappa_{ij} (a^\dagger_j a_i-a^\dagger_i a_j)
\end{equation}
is an anti-hermitian operator with orbital rotation amplitudes 
$\boldsymbol \kappa =\{\kappa_{ij}\}$. Variations of $\kappa_{ij}$ are used to
minimize the expectation value of the electronic energy
\begin{equation}
  \label{eq:E_HF}
  E_\text{HF} = \min_{\boldsymbol\kappa}\braket{\Phi(\boldsymbol
    \kappa)| H_e | \Phi(\boldsymbol \kappa)}, 
\end{equation}
where $H_e$ is the second-quantized electronic Hamiltonian,
\begin{equation}
  \label{eq:qe_ham}
  H_e = \sum_{ij} h_{ij} {a}^\dagger_i {a}_j + \frac{1}{2}\sum_{ijkl}
  g_{ijkl} {a}^\dagger_i {a}^\dagger_k {a}_l {a}_j.
\end{equation}
Here, $h_{ij}$ and $g_{ijkl}$ are one- and two-electron integrals, which depend on 
the nuclear configuration.\cite{Helgaker:2000_hf} 

If a \gls{MSO} basis $\{\psi_i\}$ contains $N_b \ge N_e$ orbitals, the
first $N_e$ orbitals are termed \emph{occupied}, while the remaining
$N_b - N_e$ of them are \emph{unoccupied} or \emph{virtual}. Only mixing
of occupied and virtual orbitals by means of elements of
$\boldsymbol \kappa$ changes the expectation value in Eq.~\eqref{eq:E_HF}.
Corresponding orbital rotations are termed \emph{non-redundant}. 
The \gls{HF} method
has polynomial complexity in the number of parameters that need to be
optimized: there are only $N_e(N_b-N_e)$ non-redundant orbital rotation
amplitudes $\{\kappa_{ij}\}$ to be determined. In most computer implementations, 
optimal parameters are found by iterative diagonalization of an $N_b \times N_b$ Fock matrix
with elements
\bea
f_{ij} = \braket{\Phi(\boldsymbol\kappa)|[\tilde{a}^\dagger_i \tilde{a}_j,H_e]|\Phi(\boldsymbol\kappa)},
\eea
which produces orbital energies $\{\epsilon_i\}$ as eignevalues 
and canonical MSOs (CMSOs) as eigenvectors for the $N_e$ system.\cite{Helgaker:2000_hf} 

The electronic Hamiltonian in \eq{eq:qe_ham} can be also written using the CMSOs 
of the $N_e$ system
\begin{equation}
  \label{eq:hec}
  \tilde{H}_e = \sum_{i} \epsilon_{i} {\tilde{a}}^\dagger_i {\tilde a}_i + \frac{1}{2}\sum_{ijkl}
  v_{ijkl} {\tilde a}^\dagger_i {\tilde a}^\dagger_k {\tilde a}_l {\tilde a}_j,
\end{equation}
where the one-electron part acquires the diagonal form. Although the \gls{HF}
energy is invariant with respect to the substitution of $H_e$ by $\tilde{H}_e$ in \eq{eq:E_HF}, 
$\tilde{H}_e$ is often used to start hybrid quantum-classical computations  
because its form is simpler and classical computers can provide the CMSOs 
relatively easily for a wide range of systems.  

\subsection{Qubit mean field}
\label{sec:qubit-mean-field-1}

Quantum computers employ two-level systems (``qubits'') as the basis
of computations,\cite{Peruzzo:2014/ncomm/4213,Kandala:2017/nature/242} 
therefore, to address the electronic structure problem 
they require transforming the fermionic Hamiltonian to an equivalent (isospectral) qubit form. 
Shorter operator expressions are obtained using the \gls{BK} transformation, and thus, we 
will employ it for transforming $H_e$ to
\begin{equation}
  \label{eq:spin_ham}
  {H}_\text{BK} = \sum_I C_I\,T_I,
\end{equation}
where $C_I$ are numerical coefficients and $T_I$ are
products of several spin operators $\omega_i^{(I)} \in \{x_i,\ y_i,\ z_i\}$ 
\begin{equation}
  \label{eq:Ti}
  T_I = \omega_{k-1}^{(I)} \cdots \omega_0^{(I)} ,\quad 1 \le k \le N_e.
\end{equation}
Here, $x_i,\ y_i,\ z_i$ are the Pauli matrices for the $i^{\rm th}$ qubit.
The maximum length of $T_I$'s determines the degree of locality
of the qubit Hamiltonian, which is $\sim\log_2 N_b$ for the \gls{BK} transformation. 
Note that a different qubit Hamiltonian is obtained if 
the electronic Hamiltonian was presented in the CMSO [\eq{eq:hec}] first and then
\gls{BK} transformed. We will denote such a Hamiltonian as $\tilde{H}_\text{BK}$.
The structure of $\tilde{H}_\text{BK}$ is similar to that of ${H}_\text{BK}$, but values of 
its coefficients $C_I$ and components of corresponding operators $T_I$ are different.  

To perform a mean-field treatment of the qubit Hamiltonians we employ the 
parametrization of the spin-1/2 Hilbert space in terms of the so-called spin coherent
states.\cite{Perelomov:1972, Radcliffe:1971/jpa/313,
  Arecchi:1972/pra/2211, Lieb:1973/cmp/327} This representation is convenient 
  because it provides a simple parametrization of the electronic energy functional.

A spin coherent state (also known as a ``Bloch state'') for a single
particle with spin $J$ is defined by the action of an appropriately scaled 
exponent of the lowering operator $\hat S_{-}$ on the normalized eigenfunction of $\hat S_z$
operator, $\hat S_z \ket{JM} = M \ket{JM}$, with maximal projection
$M = J$\cite{Lieb:1973/cmp/327}:
\begin{align}
  \label{eq:spin_coh_state}
  \ket{\Omega} = & \cos^{2J}\left(\frac{\theta}{2}\right) \exp\left[\tan
                   \left(\frac{\theta}{2}\right)\, \E^{\I\phi}\, \hat S_{-}\right]
                   \ket{JJ} \nonumber \\
  = & \sum_{M = -J}^J \binom{2J}{M+J}^{1/2} \nonumber \\
                 &\times  \cos^{J+M}{\left(\frac{\theta}{2}\right)}
                   \sin^{J-M}\left(\frac{\theta}{2}\right)
                   \E^{\I(J-M)\phi} \ket{JM},
\end{align}
where $\Omega = (\phi, \theta)$, $0 \le \phi < 2\pi$,
$0 \le \theta \le \pi$, are the spherical coordinates on a unit
sphere (``Block sphere''). States $\ket{JM}$ are normalized as:
\begin{equation}
  \label{eq:smin_act}
  \ket{JM} = \binom{2J}{M+J}^{1/2} [(J-M)!]^{-1} S_{-}^{J-M} \ket{JJ}.
\end{equation}
For the qubits used in quantum computing, $J$ is $1/2$.

 A direct product of the spin-coherent states for $N_e$ qubits,
$\ket{\boldsymbol \Omega} = \ket{\Omega_1, \Omega_2, \ldots,
  \Omega_{N_e}}$ is used as an analog of the \gls{HF} wavefunction, 
  Eq.~\eqref{eq:wf_anz}. Also, by analogy with the
\gls{HF} energy functional in the right-hand side of
Eq.~\eqref{eq:E_HF}, we introduce the \gls{QMF} energy functional as
\begin{equation}
  \label{eq:mf_func_def}
  E_\text{QMF}(\boldsymbol \Omega) = \braket{\boldsymbol \Omega |
    {H}_\text{BK} | \boldsymbol \Omega}.
\end{equation}
The explicit form of this functional can be easily obtained using the 
substitution suggested by~\citet{Lieb:1973/cmp/327} for every $T_I$ term in $H_\text{BK}$.
%\begin{equation}
%  \label{eq:letter_rep_comp}
%  \begin{array}{lcr}
%    x_i & \to & \cos\phi_i\sin\theta_i \\
%    y_i & \to & \sin\phi_i\sin\theta_i \\
%    z_i & \to & \cos\theta_i.
%  \end{array}
%\end{equation}
Minimization of the \gls{QMF} functional in \eq{eq:mf_func_def} with
respect to all Bloch angles
$(\phi_1, \theta_1, \phi_2, \theta_2, \ldots)$ gives the \gls{QMF}
energy corresponding to the electronic Hamiltonian $H_e$
\bea
E_\text{QMF} = \min_{\boldsymbol \Omega }E_\text{QMF}(\boldsymbol \Omega).
\eea
If  a different initial electronic Hamiltonian is used, for instance $\tilde{H}_e$, 
the qubit mean-field energy can potentially be different, 
$\tilde{E}_\text{QMF} = \min_{\boldsymbol \Omega }\braket{\boldsymbol \Omega |
    \tilde{H}_\text{BK} | \boldsymbol \Omega}\ne E_\text{QMF}$.  

\section{Results and discussion}
\label{sec:results-discussion}

\subsection{LiH molecule}
\label{sec:lih-molecule}

A numerical comparison of fermionic and qubit mean fields
was conducted by calculating the LiH molecule in the STO-3G basis set
at the interatomic distance $R = 1.45$ \AA~(at this distance the HF mean-field 
description is quite accurate). In this example, the electronic
structure problem has $N_b = 12$ MSOs. The electronic Hamiltonians
were considered in two sets of MSOs: 1) the \gls{HF} CMSOs 
of the LiH molecule [\eq{eq:hec}], and 2) MSOs from LiH$^{2+}$ that keep the electronic Hamiltonian 
in the form of \eq{eq:qe_ham}.
\footnote{These MSOs were obtained as the HF CMSOs for the singlet LiH$^{2+}$ cation.} 
Two qubit Hamiltonians ($\tilde{H}_\text{BK}$ and $H_\text{BK}$) were obtained 
by applying the \gls{BK} transformation to the electronic Hamiltonians.\footnote{The BK Hamiltonians 
are generated using the OpenFermion software\cite{OpenFermion:2017}.} Both qubit Hamiltonians
contain 631 $T_I$ terms, but their \gls{QMF} functionals are straightforward to formulate using 
matrix elements of the spin coherent states.\cite{Lieb:1973/cmp/327} 
%which can be found in the Supplementary Information. 
Table~\ref{tab:LiH_compreh} presents energy minima of the two \gls{QMF} functionals. 
The results indicate that \gls{QMF} is unable to recover the \gls{HF} energy if the MSOs 
used to construct the electronic Hamiltonian are not the \gls{HF} CMSOs for the state of interest.
In other words, unless the true mean-field minimum corresponding to the HF energy is 
encoded in $\tilde{H}_e$ by the choice of MSOs,  the \gls{QMF} method may not find it on its own. 

%%%%%%%%%%%%%%%%%
\begin{table}  
  \centering
  \caption{\gls{HF} and \protect\gls{QMF} energies (in a.u.) for the LiH
    molecule in the STO-3G basis at $R = 1.45$ \AA~using different sets of
    \protect\glspl{MSO}~\eqref{eq:qe_ham}. The global
    miminum of energy for this molecule in this basis set is the
    \acrfull{FCI} value, $E_\text{FCI} = -7.88098232$ a.u.}\label{tab:LiH_compreh}
      \begin{ruledtabular}
    \begin{tabular}{@{}lc@{}}
    $E_\text{QMF}$ (MSOs of LiH$^{2+}$) & -7.77340369 \\
    $\tilde{E}_\text{QMF}$ (HF CMSOs of LiH) & -7.86256779 \\
    $E_\text{HF}$ & -7.86256779 
    \end{tabular}
  \end{ruledtabular}
\end{table}

%%%%%%%%%%%%%%%%%%%%%%%%%%%%%%%%%%%%%%%%%%%%%%%%%%%%%
\subsection{One-particle Hamiltonians}
\label{sec:model-indep-electr}

For a better understanding of the QMF shortcomings illustrated in the LiH case, we 
consider a sum of one-electron operators constituting the  
Fock operator for the first $n = 8$ (out the total $N_b = 12$) CMSOs of the LiH molecule at 
$R(\ce{Li-H}) = 1.45$ \AA 
\begin{equation}
  \label{eq:He_1p}
  \tilde{H}_e^{(1)} = \sum_{i = 1}^n \epsilon_i \tilde{a}_{i}^\dagger\tilde{a}_{i} %+ \tilde{a}_{2i}^\dagger \tilde{a}_{2i}\right),
\end{equation}
where $\epsilon_i$'s are $\{{-2.3462},\, {-0.2975},\, {0.0793},\,
{0.1635}\}$, note that all orbital energies are doubly degenerate due to the spin symmetry. 
Considering these orbital energies, it is easy to guess that the global
minimum of the electronic energy for $\tilde{H}_e^{(1)}$ corresponds to $N_e = 4$
electrons (as in neutral \ce{LiH}) and equals to
$E_0 = \sum_{i=1}^{N_e}\epsilon_i = {-5.2874}$ a.u. After the \gls{BK}
transformation, $\tilde{H}_e^{(1)}$ becomes
\begin{align}
  \label{eq:H_1p_BK}
  & \tilde{H}_\text{BK}^{(1)}  = -{2.4009} +{1.1731}z_0(1+z_1) \nonumber \\
  &        - 0.03965z_4(1+z_5) - {0.08175}z_6(1+z_7z_5z_3) \nonumber \\
  &  + {0.14875}z_2(1+z_3z_1) .
\end{align}
Clearly, this Hamiltonian can be solved exactly using \gls{QMF}  because 
it only contains commuting $z_i$ operators and its eigenfunctions are products 
of $z_i$ eigenfunctions. This property is general for the \gls{BK} transformation: any
$\tilde{a}^\dagger_i \tilde{a}_i$ term results in a product of $z_i$ operators, see Eq.~45 in
Ref.~\citenum{Seeley:2012/jcp/224109}. 
Converting $\tilde{H}_\text{BK}^{(1)}$ into the \gls{QMF} functional %by the rule~\eqref{eq:letter_rep_comp} 
and finding its minimum gives $\tilde{E}_\text{QMF}^{(1)} = E_0$.  

Let us perform a unitary transformation [\emph{cf.} Eq.~\eqref{eq:Ua}] within 
the occupied-virtual subspace 
\begin{align}
  \label{eq:orb_transf}
  b_{2}  = & \phantom{-}\cos\tau\, \tilde{a}_{2} + \sin\tau\, \tilde{a}_{3},
                     \nonumber \\
  b_{3}  = & -\sin\tau\, \tilde{a}_{2} + \cos\tau\, \tilde{a}_{3},
\end{align}
the creation counterparts (Hermite conjugates) of $\tilde{a}_{2}$ and $\tilde{a}_{3}$ 
 are transformed accordingly, and all the
remaining operators are left untouched.\footnote{The list of the MSOs is enumerated 
so that all spin $z$-projection $+1/2$ orbits in the ascending order of the orbital energies 
are followed by the corresponding spin $z$-projection $-1/2$ subset.} The new fermionic operators
$(b_i^\dagger,\, b_i)$ satisfy the same commutation relations as the
old ones. Of course, this transformation can be easily undone by
taking $\tau' = -\tau$, and hence, the \gls{HF} method applied to the transformed 
fermionic Hamiltonian will recover the original value for the minimum.
However, the rotated electronic Hamiltonian defines a new \gls{BK}
image, which for the particular choice $\tau = \pi/6$ reads
\begin{align}
  \label{eq:H_1p_BK_rot}
  & H_\text{BK}^{(1)} = -{2.4009} - {0.08175}z_6(1+ z_7z_5z_3) \nonumber \\
  & + {1.1731} z_0 (1+ z_1) +{0.14875}\,z_2 - {0.03965}\, z_4 \nonumber \\
  & + {0.10165}\,   z_3z_2z_1 + {0.00745}\,   z_5z_4 \nonumber \\
  & - {0.0815796}(x_5z_4x_3  +y_5y_3z_2z_1). 
\end{align}
Here, the pairs of $x_i$ and $y_i$ operators introduce entanglement in 
eigenstates of the rotated Hamiltonian and make the \gls{QMF} ground state 
solution higher in energy than the exact one:
$E_\text{QMF}^{(1)} = -5.2081$ a.u. $> E_0$
% and corresponds to $N = 5$ electrons, in other words to an anionic
% state! A minimum for $N = 4$ electrons (obtained with the particle
% number constraint) lies higher at {-5.1932}.

Further testing with random unitary transformations involving more MSOs than in \eq{eq:orb_transf} 
showed that the \gls{QMF} minima are always higher than $E_0$ as long as the
occupied-virtual rotations are included in the unitary transformation.
% Some of them are even characterized by fractional $N$.
Thus, the source of the \gls{QMF} problems can be related to its inability to perform 
rotations equivalent to the unitary one-particle transformations that are easily achieved by 
the one-electron operators in \eq{eq:Ua}. 
%method cannot reach
%the minimum of the fermionic independent-particle Hamiltonian.

The last example can make an impression that the fermionic mean-field approach is more powerful 
than its qubit counterpart. The following example shows that it is not the case.  
A simple qubit Hamiltonian that allows for the exact solution by \gls{QMF} 
can be written as 
\begin{equation}
  \label{eq:qubit_simple}
  H_q^{(1)} = \alpha z_0 + \beta x_1 + \gamma y_2,
\end{equation}
it can be converted to a fermionic form using the inverse of the \gls{JW}
transformation (see Ref.~\citenum{Nielsen:2005/scholar_text}):
\begin{align}
  \label{eq:qubit_simple_ferm}
  H_e^{(1q)} = & \alpha (1 - 2a_1^\dagger a_1) + \beta (1 -
                  2a_1^\dagger a_1)(a_2^\dagger + a_2) \nonumber \\
                & + \I\gamma (1 - 2a_1^\dagger a_1)(1 -
                  2a_2^\dagger a_2)(a_3^\dagger - a_3).
\end{align}
It is easy to see that the $H_e^{(1q)}$ Hamiltonian is \emph{quintic} in terms of
powers of elementary fermionic operators, and it does not conserve
the number of electrons. Thus, any \gls{HF} procedure that uses Slater 
determinants with a fixed number of electrons will give the minimum energy that is 
higher than the exact one.

\subsection{Symmetry breaking}
\label{sec:when-does}

\begin{figure}
  \centering %
  \includegraphics[width=1.0\columnwidth]{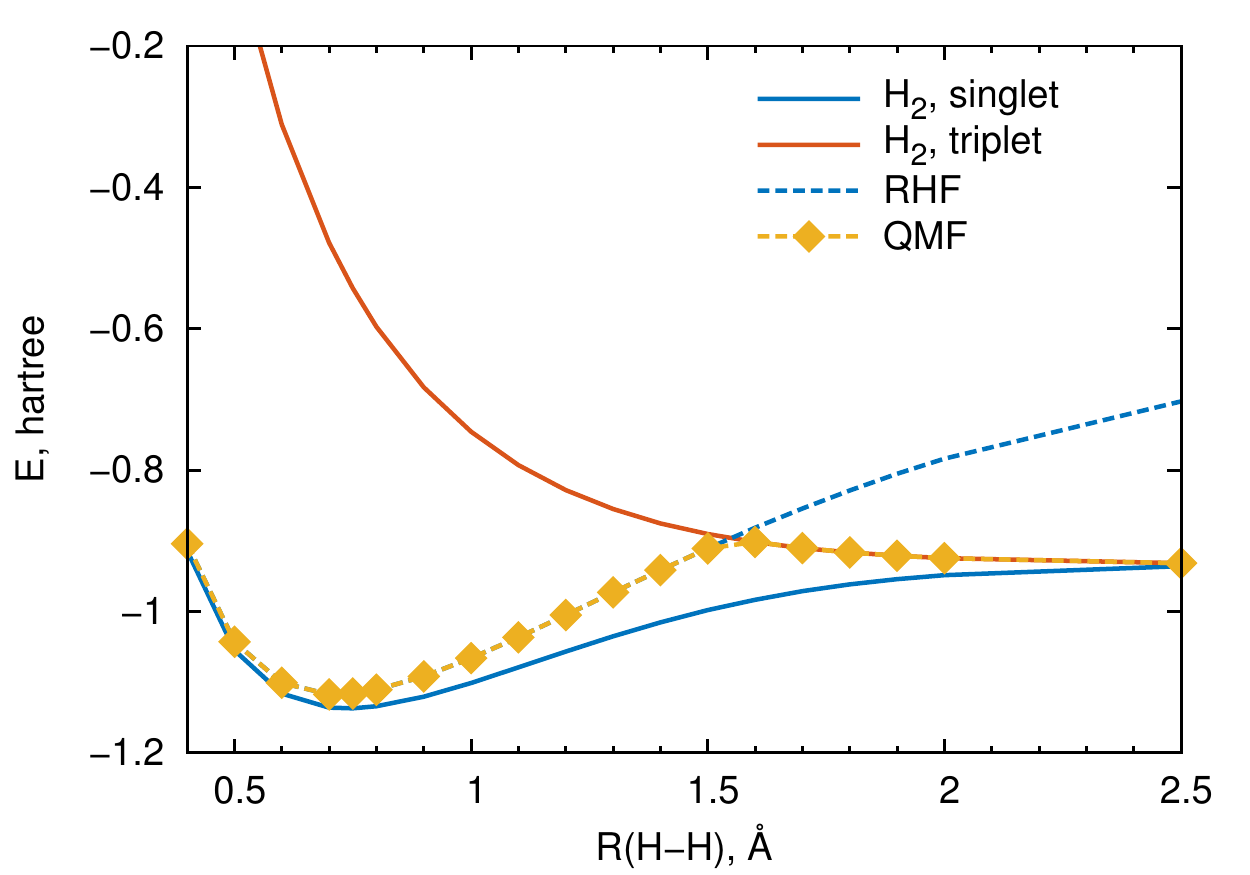}
  \caption{Two lowest eigenstates of the
    Hamiltonian $H_{\rm BK}(R)$ (solid lines), and the PES 
    corresponding to the minimum of the \protect\gls{QMF} functional,
    Eq.~\eqref{eq:mf_func_def} (dashed line).}
  \label{fig:h2_exact_vs_qmf_unconstr}
\end{figure}

If electron correlation is large, which is usually the
case when the system ground state becomes quasi-degenerate, it is possible
to have $E_\text{QMF} \le E_\text{HF}$. A specific example is 
a stretched \ce{H_2} molecule in the STO-3G basis 
(Fig.~\ref{fig:h2_exact_vs_qmf_unconstr}). 
For $\text{R(H-H)} \gtrapprox 1.5$ \AA~the \gls{QMF} energy
plunges below the restricted (singlet) \gls{HF} energy and the
potential energy curve in this region displays a ``kink''.
This takes place because the energy of a \emph{triplet} state becomes
lower than the \gls{HF} singlet and \gls{QMF} readily switches to
the triplet solution. This phenomenon is known as the symmetry
breaking, and \gls{QMF} is prone to it: \gls{QMF} solutions may easily
violate such physical constraints as the number of electrons in the system
or the state multiplicity as in the given example.\cite{Ryabinkin:2018/unpub}
On the other hand, the fermionic mean-field approach is organized so that the unitary 
rotations [\eq{eq:Ua}] commute with the operators of the number of electrons, 
the total electron spin $z$-projection, and even with the square of the total spin 
$\hat S^2$ in some cases. 
Therefore, the fermionic mean-field approach conserves many symmetries by 
construction.\cite{JimenezHoyos:2012ep}

\section{Conclusions}
\label{sec:conclusions}

We introduced and studied properties of the \gls{QMF} approach, 
which is the simplest parametrized molecular
wavefunction ansatz possible on a quantum computer. We considered the
representation of quantum states of individual qubits in terms of the
spin coherent states that depend on a set of Bloch angles. This
representation leads to extremely simple form of the energy functional
and allows us to draw parallels between the fermionic
mean-field approach (the restricted \gls{HF} method) and \gls{QMF}.
We compared the results of \gls{HF} energy minimization with
those given by \gls{QMF} and found that generally, the
\gls{QMF} method provides energies that are higher than their
restricted \gls{HF} counterparts. This result emphasizes the importance of specifying 
the electronic Hamiltonian representation (the \gls{MSO} set) used 
for any \gls{VQE} simulations. 
Also, our observations suggest that prior \gls{HF} calculations on a 
\emph{classical} computer may be beneficial to decrease 
computational burden placed on a quantum
computer. However, this supremacy of fermionic mean-field stems from  
a somewhat biased starting point of the electronic Hamiltonian, which is formulated using 
the fermionic operators. It is easy to find an oppositely biased starting point of a spin 
Hamiltonian that can be treated exactly with 
\gls{QMF} but whose treatment with the HF method will be only approximate.

Another significant aspect of the relation between qubit and fermionic 
mean-fields is their behavior in strongly correlated cases 
where mean-field solutions become inaccurate. 
The fermionic mean-field approach usually conserves many symmetries by construction,  
and the symmetry breaking requires relaxing some constraints built in 
its variational ansatz. \gls{QMF} is prone to breaking physical symmetries because arbitrary 
qubit rotations are generally incompatible with symmetries of the electronic Hamiltonian. 
In the symmetry breaking case, PESs exhibit characteristic ``kinks'' due to drastic change
of the physical nature of the wavefunction. Although in such cases 
$E_\text{QMF} \le E_\text{HF}$, one should prefer to stay on the
same physical branch of a solution. A treatment producing continuous 
PESs in symmetry breaking situations is proposed in Ref.~\citenum{Ryabinkin:2018/unpub}, and the
current work  emphasizes additionally the importance of physical
constraints for variational methods working with qubits.

\subsection{Acknowledgements}

The authors thank P. Brumer for stimulating discussions.
A.F.I. acknowledges financial support from Natural Sciences and 
Engineering Research Council of Canada (NSERC) through the Engage grant.

%\bibliography{qcomp-snap,books,programs}%,qmfloc}

%

\end{document}